\documentclass[fleqn,10pt]{wlscirep}
\usepackage[utf8]{inputenc}
\usepackage[T1]{fontenc}
\usepackage{makecell}
\usepackage{multirow}
\usepackage{rotating}
\usepackage{colortbl}

\title{A Comprehensive Approach to Characterize Navigation Instruments for Magnetic Guidance in Biological Systems}

\author[1,*]{Peter Blümler}
\author[2,3,4*]{Fabian Raudzus}
\author[1]{Friederike Schmid}

\affil[1]{University of Mainz, Institute of Physics, 55128 Mainz, Germany}
\affil[2]{Kyoto University, Department of Clinical Application, Center for iPS Cell Research and Application (CiRA), Kyoto, Japan}
\affil[3]{Kyoto University, Neuronal Signaling and Regeneration Unit, Graduate School of Medicine, Kyoto, Japan}
\affil[4]{Kyoto University, Medical Education Center/International Education Section, Graduate School of Medicine, Kyoto, Japan}
\affil[*]{bluemler@uni-mainz.de, fabian.raudzus@cira.kyoto-u.ac.jp}


\begin{abstract}
The non-invasive spatiotemporal control of cellular functions, organization of tissues, and even the behavior of small animals has become paramount for advanced therapies. As magnetic fields do not interact with biological matter, their application is not only suitable for in vitro experiments but also for in vivo applications, even in deep tissues. Particularly, the remote manipulation of paramagnetic entities through magnetic instruments has emerged as a promising approach across various biological contexts.\\
Despite similarities in basic experimental concepts, variations in the properties and descriptions of those magnetic instruments among the authors and studies resulted in a lack of reproducibility and comparability. Therefore, this article addresses the question of how to standardize the characterization of magnetic instruments.\\
Our emphasis lies on the ability of magnetic systems to control the movement of paramagnetic objects such as ferro- or superparamagnetic particles, within organisms. This movement is achieved by exerting a force on magnetic particles by exposing them to a locally varying magnetic field. While it is well-known that the exerted force depends on the spatial variation (i.e. the gradient) of the magnetic field, the magnitude of the field is equally important. However, this second factor is often neglected in the literature. Therefore, we conduct a comprehensive analysis and discussion of both factors. Furthermore, we propose a novel descriptor, termed ‘effective gradient’, which combines both dependencies. To illustrate, we characterize different magnet systems by calculating and comparing the different quantities and relating them to two experiments with different superparamagnetic nanoparticles.
\end{abstract}

\begin{document}
\flushbottom
\maketitle
\thispagestyle{empty}
\section*{Introduction}
For research, diagnosis, and the application of novel therapies it is often necessary to spatiotemporally control cellular functions, the organization of tissues, or the location of substances remotely. Within living organisms, achieving these goals with minimal interaction with the biological system presents a significant challenge. Unfortunately, common techniques utilizing light or ultrasound face limitations due to the energy transfer to the cells, and depth of tissue penetration. Conversely, techniques based on magnetic fields present an attractive alternative that is gaining popularity. As most biological substances such as organic materials and tissues are diamagnetic or only slightly paramagnetic, they interact very weakly with magnetic fields. Consequently, biological substances can be penetrated non-invasively, and the depth of penetration is just limited by the decay of the magnetic field over distance. On the other hand, objects made from strongly paramagnetic (i.e. superpara-, ferro- or ferrimagnetic) materials, undergo significant magnetization in the presence of magnetic fields. This enables their manipulation by magnetic forces, facilitating remote movement or actuation within larger diamagnetic environments.\\
This fundamental principle is currently harnessed for various biological applications. One notable application is the magnetic guidance of objects through living organisms for diagnostic or therapeutic purposes. This concept can be applied to objects spanning a wide range of sizes, from millimeters (e.g. endoscopes and capsules\cite{Munoz_2018, Swain_2008}), micrometers (e.g. microrobots\cite{Sun_2022, Koleoso_2020, Yang_2020, Shao_2021}) to nanometers (nanoparticles). The latter are employed in drug delivery (magnetic drug delivery or targeting \cite{Shapiro_2015, Tietze_2015, Attia_2019}), tissue manipulation (by actuation\cite{Dobson_2018}, hyperthermia\cite{Rajan_2020,Obaidat_2015}, or other localized therapies\cite{Day_2021,Chansoria_2023}) at target sites. Furthermore, magnetic guidance can be utilized to move cells\cite{Bongaerts_2020, Plen_2022}, or cellular extensions such as neurites\cite{De_Vincentiis_2020,Dhillon_2022}, for applications in tissue engineering and regenerative medicine\cite{Parfenov_2020, Lee_2014, Friedrich_2021, Goranov_2020}. In this context, the strategy involves incorporating nanoparticles into the cells and subsequently applying force through the particles within the cells. Typically, superparamagnetic nanoparticles are used, often coated to improve their biocompatibility, and may have other functional groups or payloads\cite{Hauser_2015}. Moreover, protein-functionalized superparamagnetic nanoparticles demonstrate the capacity to modulate intracellular signaling\cite{Etoc_2013, Raudzus_2020}, or activate cell receptors and ion channels\cite{Lee_2021}. Magnetic guiding can also be used in microfluidics and nanomechanics\cite{Zahn_2001}. For recent reviews over this very active and interdisciplinary field, we refer to references\cite{Panina_2022, Garello_2022, Andrae_2007}.\\
In all of these applications, the generation of a suitable magnetic field either by resistive or superconducting electro-, or permanent magnets, is a prerequisite for precise steering, guiding, or actuation. Various guiding concepts\cite{Bluemler_2021, Cao_2020, Komaee_2012} and instruments\cite{Sliker_2015, Liu_2019} have been proposed, leading to a plethora of designs. Comparing such instruments is challenging due to the fact that two distinct factors contribute to the efficiency of magnetic guidance. The first is the spatial change of the magnetic field, the so-called gradient, which predominantly determines the force exerted on a particle carrying an oriented magnetic dipole. The second is the absolute (local) strength of the magnetic field, required to induce a dipole in the particle or, in the case particles with a permanent dipole, to orient them. Both factors must be jointly considered to predict the resulting magnetic force. Unfortunately, the second factor tends to be overlooked in literature, which results in confusing and misleading discussions of magnetic designs. 
The publication aims to quantitatively elucidate the interplay between strength and gradient of magnetic fields in generating the magnetic force on particles. The full mathematical formalism involves a tensorial description of magnetic gradients, which complicates the direct comparison of designs. As a practical tool to compare the efficiency of different magnet types for magnetic guiding, we propose a modified physical quantity. This quantity can be conceptualized as an effective magnetic field gradient, presented as a simple three- dimensional vector instead of a tensor. We discuss the suitability of this effective gradient is discussed for two commonly used magnet geometries in magnetic guiding.
\begin{table}[h!]
\centering
\renewcommand{\arraystretch}{2}
\begin{tabular}{c|l|c|l}
symbol & terminology & units & conversion \\
\hline\hline
$\vec{B}$ & \makecell[l]{magnetic flux density, \\ magnetic induction} & \makecell[l]{T = Vs/m \\ \;\;\;= kg/(As\textsuperscript{2})} & $\vec{B} = \mu_0 (\vec{H} + \vec{M}) = \mu \vec{H}$ \\
\hline
$\chi$ & magnetic susceptibility &  & $\chi = M/H$ \\
\hline
$\vec{H}$ & magnetic field, magnetizing field & A/m &  \\
\hline
$k_\text{B}$ & Boltzmann constant & J/K &  $k_\text{B} \equiv 1.380649 \cdot 10^{-23}$ kg\,m\textsuperscript{2}/(K\,s\textsuperscript{2})\\
\hline
$\mu$ & permeability & Vs/(Am) &  $\mu = \mu_0 (1 + \chi)$\\
\hline
$\mu_0$ & permeability of vacuum & Vs/(Am) & \makecell[l]{$\mu_0 \equiv 1.25663706212 \cdot 10^{-6}$ kg\,m/(As)\textsuperscript{2} \\ \quad\, $\approx 4\,\pi \cdot 10^{-7}$ kg\,m/(As)\textsuperscript{2}}\\
\hline
$\vec{m}$ & magnetic moment & Am\textsuperscript{2} & $\vec{m} = \iiint \vec{M} \text{d}V \approx \;M_\text{s}V$\\
\hline
$\vec{M}$ & magnetization & A/m & $\vec{M} = \text{d}\vec{m}/\text{d}V \approx \; \Vec{m}/V$\\
\hline
$\vec{\nabla}$ & spatial gradient & 1/m & $\vec{\nabla} = (\partial/\partial x,\;\partial/\partial y,\;\partial/\partial z)$\\
\hline
$r$ & radius (of a particle) & m & \\
\hline
$T$ & absolute temperature & K & \\
\hline
$V$ & volume & m\textsuperscript{3} & sphere: $V = 4/3 \pi r^3$\\
\hline
\end{tabular}
\caption{Used symbols and corresponding SI unit in this manuscript. It has to be mentioned that magnetic field, $H$, and magnetic flux density, $B$, are used synonymously in the text to adapt to common nomenclature of most publications. When not specifically mentioned this refers to magnetic flux density.}
\label{tab1}
\end{table}
\section*{Theory}
\subsection*{Average magnetization of superparamagnetic nanoparticles}
At room temperature, a ferro- or ferrimagnetic material consists of small volume regions known as Weiss domains, wherein the magnetic dipole moments of strength $\vec{m}_i$, are mutually aligned in the same direction. The density of the sum of all magnetic moments yields the magnetization (see Tab. \ref{tab1})
\begin{equation}\label{eq:1}
\vec{M} = \frac{1}{V} \sum_i \vec{m}_i . \\
\end{equation}
These domains are microscopic in size and separated by (Bloch-) walls. Larger objects consist of numerous such domains (see Fig. 1) with different magnetization directions. The total magnetization $\langle \vec{M} \rangle$ of the object is the vector sum of the magnetizations of all domains weighted by their volume. Figure 1 shows how an external magnetic field induces shifts of the domain walls, thereby enlarging domains with magnetization components in the same direction as the magnetic field.\\
This process leads to an increase in total magnetization from zero ($\langle \vec{M}(\vec{B}=0) \rangle$) in the absence of a magnetic field -- where all subdomain magnetizations cancel each other -- to maximum (saturation) magnetization ($\langle |\vec{M}| \rangle = M_\text{s}$) at high magnetic fields. Each individual domain is fully magnetized ($m_i = M_\text{s}V_i$ where $V_i$ is the size of the domain), but their directional average $\langle \vec{M}(\vec{B}) \rangle$ depends on the magnetic field.\\ 
When the size of the object is of the order of such a domain (typically $r < 30$~nm), the object eventually consists of only one single, fully magnetized domain. However, due to their small size, their rotational Brownian motion will have a similar effect on the thermally averaged magnetization, $\langle \vec{M}(\vec{B}) \rangle$, as illustrated on the right side of Fig. 1. The resulting value can be approximated by a Langevin-function, $\mathfrak{L}(x)$ \cite{Knopp_2012,Krishnan_2016}: The average magnetization $\langle \vec{M} \rangle$ points in the direction of $\vec{B}$, and its amplitude is given by
\begin{equation}\label{eq:2}
\begin{split}
   &\big|\langle \vec{M}(\vec{B}) \rangle \big| \;\; = \;\; \frac{m}{V}\, \mathfrak{L}\left(\frac{mB}{k_\text{B}T}\right) \;\; \approx \;\; M_\text{s} \mathfrak{L}\left(\frac{M_\text{s}VB}{k_\text{B}T}\right)   \\[2ex]
   &\text{where} \quad \mathfrak{L}(\xi) \approx \coth{\xi}-\frac{1}{\xi} \quad \text{and} \quad \xi = \frac{M_\text{s}V}{k_\text{B}T} B \quad \text{with} \quad B = \sqrt{B_x^2 + B_y^2 + B_z^2} .
\end{split}
\end{equation}
While each particle is fully magnetized, its time-averaged magnetization follows eq. (\ref{eq:2}). This behavior is known as superparamagnetism\cite{Bean_1959}. Chemical stabilization is required to avoid agglomeration due to mutual dipolar attraction of the particles\cite{Kemgang_2022}. Superparamagnetic particles form the basis of ferrofluids\cite{Odenbach_2009} and find numerous applications in theranostics\cite{Caizer_2022}. Independent of their size and ignoring hysteresis effects in the case of multi-domain particles, the macroscopic magnetization  $\langle \vec{M}(\vec{B}) \rangle$ depends on the field according to eq. (\ref{eq:2}) and is also strongly dependent on $M_\text{s}$ and the size/volume $V$ of the magnetic particle (MP).\\
This is illustrated by Fig. \ref{fig2} which demonstrates that the field dependence of $\langle \vec{M}(\vec{B}) \rangle$ or the Langevin-function is linear for very small fields and its steepness strongly depends on the particle size. The constant saturated region where $M \approx M_\text{s}$ is reached at lower magnetic fields for larger particles. At higher magnetic fields, the behavior of small particles varies more strongly as a function of field strength than that of larger particles. This also implies that in the same magnetic instruments, smaller particles exhibit a stronger field dependence. 
We emphasize that eq. (\ref{eq:2}) relates the \textit{amplitudes} of $\vec{M}, \vec{m}, \text{and } \vec{B}$ to each other, despite their status as spatial vectors. As noted above, $\langle \vec{M} \rangle$ is parallel to $\vec{B}$. Given that MPs carry a permanent magnetization, the magnetic field induces a torque, causing a rotation of the whole particles to align their magnetic moments with $\vec{B}$. As a consequence, it can be assumed that (with $\cdot$ being the dot or scalar product of two vectors)
\begin{figure*}[t!]
\centering
\includegraphics[width=0.6\textwidth]{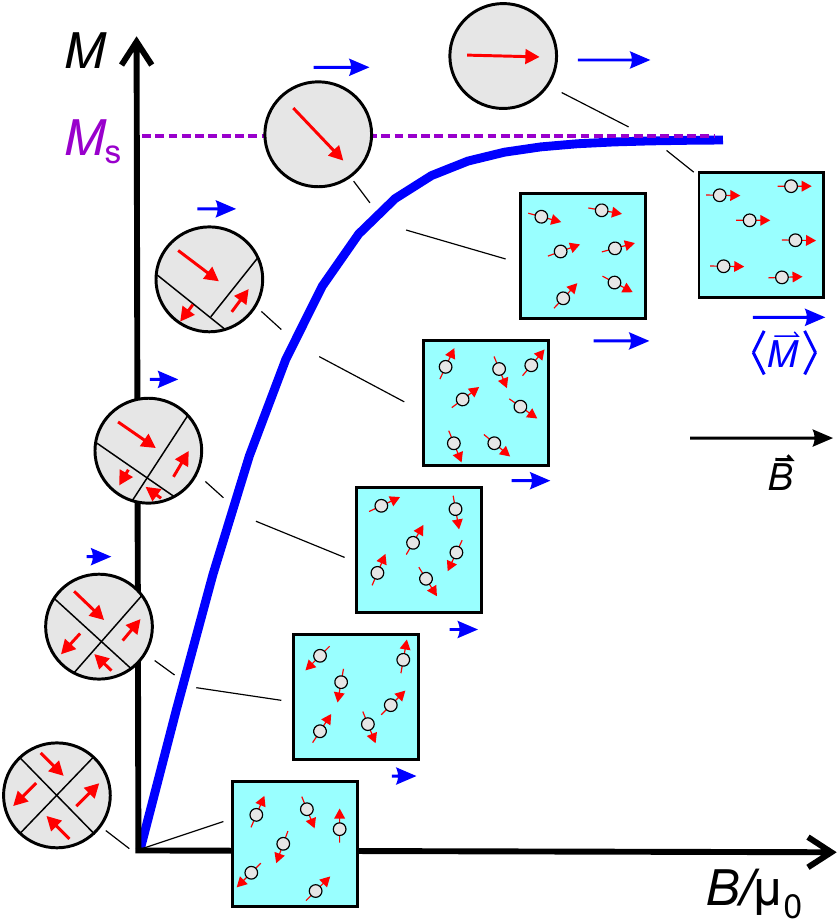}
\caption{Schematic illustration of the magnetization process of large particles with several domains (gray circles left of curve) and superparamagnetic nanoparticles (light blue squares right of curve). The thick blue line shows the average magnetization as a function of applied magnetic field, which points to the right. The curve corresponds to the positive quadrant of Fig. \ref{fig2}. Six snapshots illustrate the growth of domains or particle orientation along the field direction. The resulting averaged magnetization $\langle \vec{M}(\vec{B}) \rangle$ is sketched by the blue arrows, their lengths correspond to the value on the ordinate. The red arrows indicate the magnetization of each domain or particle.}
\label{fig1}
\end{figure*}
\begin{equation}\label{eq:3}
\langle \vec{M} \rangle \parallel \vec{B} \quad \text{or} \quad \langle \vec{M} \rangle \cdot \vec{B} \; =  \; \big| \langle \vec{M} \rangle \big| \, \big| \vec{B} \big| \;\; \approx \;\; M_\text{s}\, \mathfrak{L}(\xi)\, B , \\
\end{equation}
\noindent where $\xi$ is defined as in eq. (\ref{eq:2}).
\begin{figure*}[t!]
\centering
\includegraphics[width=0.4\textwidth]{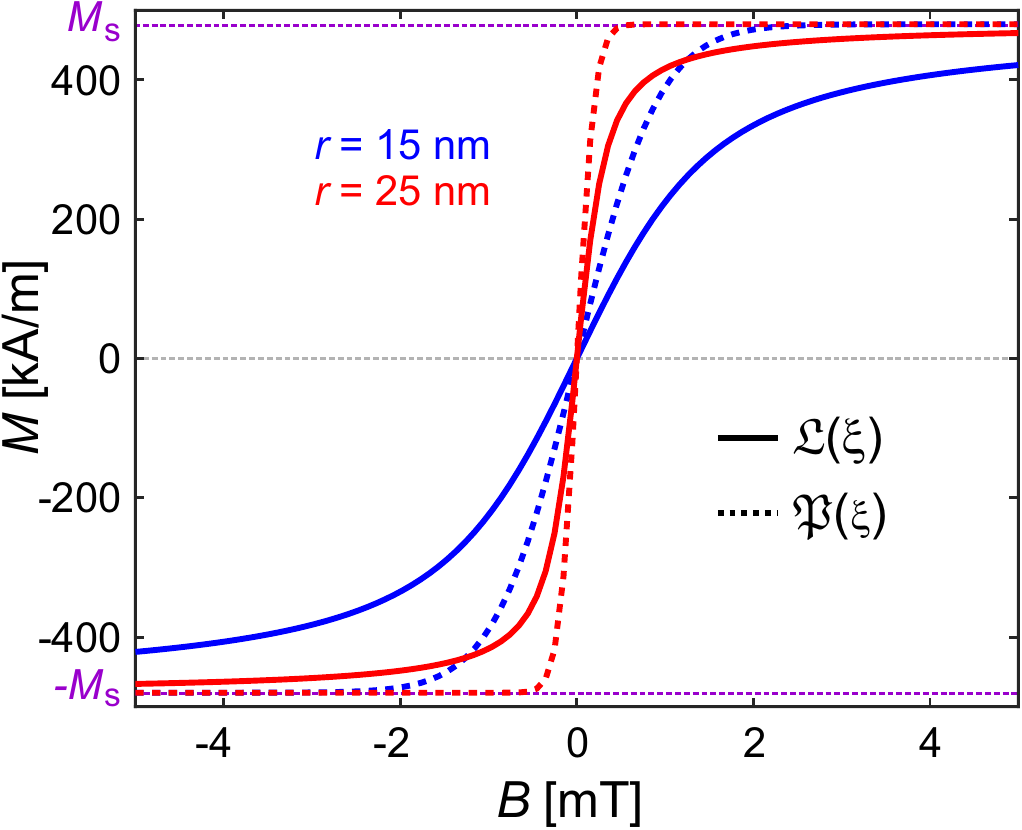}
\caption{Field dependence of the magnetization $\langle \vec{M}(\vec{B}) \rangle$  of superparamagnetic iron-oxide MPs with two different sizes $r = 15$~nm (blue) and $r = 25$~nm (red) with $M_\text{s} = 480$ kA/m at $T = 300$~K (values from\cite{Bender_2018}). The solid lines show $\langle \vec{M}(\vec{B}) \rangle$ as approximated by eq. (\ref{eq:2}) or $\mathfrak{L}(\xi)$. The dashed lines show $\mathfrak{P}(\xi)$ in comparison. ($\mathfrak{L}(\xi)$ is defined in eq. (\ref{eq:9}) and describes the amplitude of the magnetic force as a function of applied field for a given magnetic gradient).}
\label{fig2}
\end{figure*}
\subsection*{The magnetic field and its gradient}
Another ingredient for calculating the force is the spatial variation of the external magnetic field. It is quite intuitive that magnetized MPs experience no drag in a perfectly homogeneous magnetic field, much like a ball on a perfectly flat plate, as moving the particle to different position changes nothing. Drag forces only arise when there is a change in potential energy with respect to space. In the example of the ball, the potential energy is proportional to its height, and a force can be induced by tilting the plane, such that the height depends on the lateral position. Analogously, the theory of electromagnetism tells us that we can assign a ‘magnetic potential energy’ to a particle with magnetic moment $\vec{m}$ in a magnetic field $\vec{B}$, given by $\vec{E} = \vec{B}\vec{m}$\,\cite{Jackson_1975}. This allows us to induce a magnetic force on the MPs by spatial variations in the magnetic field,  $\vec{B}$. However, the situation is complicated by the fact that the magnetic field has three components that can all change with space. The spatial variation must be described in terms of a second rank tensor (with $\otimes$ being the tensor product)
\begin{equation}\label{eq:4}
\overline{\overline{G}} \;\; = \;\; \mathbf{grad}\,\vec{B} \;\; = \;\;  \vec{\nabla}\,\otimes\,\vec{B} \;\; = \;\; \left( \begin{array}{c c c}
\frac{\partial{B_x}}{\partial{x}} & \frac{\partial{B_x}}{\partial{y}} & \frac{\partial{B_x}}{\partial{z}}\\[1ex]
\frac{\partial{B_y}}{\partial{x}} & \frac{\partial{B_y}}{\partial{y}} & \frac{\partial{B_y}}{\partial{z}}\\[1ex]
\frac{\partial{B_z}}{\partial{x}} & \frac{\partial{B_z}}{\partial{y}} & \frac{\partial{B_z}}{\partial{z}}\\ \end{array} \right)
\end{equation}
\noindent where $\partial{B_i}/\partial{x}$  etc. are partial derivatives. Fortunately, as we will soon see, the full knowledge of the tensor is not necessary when calculating the magnetic force on MPs. Nevertheless, eq. (\ref{eq:4}) demonstrates that a magnetic system cannot be described by a single value for a magnetic field gradient. Indeed, it is not even possible to generate a single gradient outside magnets (such as, e.g., ($\partial{B_x}/\partial{x}  \neq 0$ , with all other partial derivatives in $\mathbf{grad}\,\vec{B}$ being zero). This is forbidden by the Maxwell equations for electromagnetism, i.e. (with $\times$ as the cross product of two vectors) 
\begin{equation}\label{eq:5}
\begin{split}
   \mathbf{div}\,\vec{B} \;\; &= \;\; \vec{\nabla}\,\cdot\,\vec{B}\;\; = \;\;\frac{\partial{B_x}}{\partial{x}}+\frac{\partial{B_y}}{\partial{y}}+\frac{\partial{B_z}}{\partial{z}}\;\; &= \;\; 0,\\[3ex]
   \mathbf{rot}\,\vec{B} \;\; &= \;\;\vec{\nabla}\,\times\,\vec{B}\;\; = \;\;\left( \begin{array}{c}
\frac{\partial{B_z}}{\partial{y}} - \frac{\partial{B_y}}{\partial{z}}\\[1ex]
\frac{\partial{B_x}}{\partial{z}} - \frac{\partial{B_z}}{\partial{x}}\\[1ex]
\frac{\partial{B_y}}{\partial{x}} - \frac{\partial{B_x}}{\partial{y}}\\
\end{array} \right) \;\; &= \;\; 0.
\end{split}
\end{equation}
Thus, each gradient must be accompanied by at least another gradient of equal strength and opposite direction or sign. A single magnetic field gradient cannot exist!
\subsection*{Magnetic force}
As discussed in the previous section, the force on a MP is given by the gradient of the magnetic potential energy. In the context of an MP with magnetic moment $\vec{m}$ subjected to a magnetic field of strength $\vec{B}$ , this potential energy is given by\cite{Jackson_1975}.
\begin{equation}\label{eq:6}
\vec{F}_\text{m} \;\; = \;\; \vec{\nabla} (\vec{m}\,\cdot\,\vec{B}) . \\
\end{equation}
Averaging over the distribution of $\vec{m}$ and using eq. (\ref{eq:2}), one obtains
\begin{equation}\label{eq:7}
\langle \vec{F}_\text{m}\rangle \;\; = \;\; \vec{\nabla} (V\langle\vec{M}\rangle\,\cdot\,\vec{B}) . \\
\end{equation}
From eq. (\ref{eq:3}) it is known that $\langle \vec{M} \rangle$ and $\vec{B}$ are parallel. This presents a challenge in magnetic guiding, as the dot product cancels the sign of the magnetic field, causing MPs to be constantly driven in the direction of regions with higher magnetic fields. This is completely analogous to paper clips being equally attracted by the south and north pole of a permanent magnet (never repelled!). However, it allows us to re-express eq. (\ref{eq:7}) as (here $\xi = \frac{M_\text{s}V}{k_\text{B}T}\,B$ is defined as in eq. (\ref{eq:2}))
\begin{equation}\label{eq:8}
\begin{split}
   \langle \vec{F}_\text{m}\rangle \;\; &= \;\;\vec{\nabla} \left( M_\text{s}V\,\mathfrak{L}(\xi)\,B \right) \;\; = \;\;  M_\text{s}V\,\mathfrak{L}(\xi)\,\vec{\nabla}B\;+\; BM_\text{s}V\;  \vec{\nabla}\mathfrak{L}(\xi) \\[2ex]
   &= \;\;\left( M_\text{s}V\,\mathfrak{L}(\xi)\;+\; B\,\frac{\text{d}\mathfrak{L}}{\text{d}\xi}\,k_\text{B}T \right) \vec{\nabla}B \;\; = \;\; M_\text{s}V \left( \mathfrak{L}(\xi)\;+\;\xi\, \frac{\text{d}\mathfrak{L}}{\text{d}\xi} \right) \vec{\nabla}B \\
\end{split}
\end{equation}
\begin{equation}\label{eq:9}
\begin{split}
   \quad \quad \;&=: \;\;M_\text{s}V\;\mathfrak{P}(\xi) \left( \tfrac{\partial{B}}{\partial{x}},\; \tfrac{\partial{B}}{\partial{y}},\;\tfrac{\partial{B}}{\partial{z}} \right)  \;\; = \;\; k_\text{B}T  \;\mathfrak{P}(\xi) \left( \tfrac{\partial{\xi}}{\partial{x}}, \;\tfrac{\partial{\xi}}{\partial{y}},\;\tfrac{\partial{\xi}}{\partial{z}} \right) \\[2ex]
    \text{with} &\quad \mathfrak{P}(\xi)\;:=\; \coth(\xi)\;-\;\xi \text{csch}^2(\xi) \quad \text{and} \quad B\;=\;|\vec{B}|\;=\;\sqrt{B^2_x+B^2_y+B^2_z}, \\
\end{split}
\end{equation}
Compared to eq. (\ref{eq:6}), this equation is much simpler, as it no longer requires the calculation of the complicated gradient tensor (as seen in eq. (\ref{eq:4})). Due to the loss of direction in the dot product of eq. (\ref{eq:3}), the magnetic force now only depends on the rescaled strength of the magnetic field, $\xi$, and its gradient $\vec{\nabla}\xi$, a simple vector. However, we have introduced a new non-linear, material- and field-dependent function $\mathfrak{P}(\xi)$, which accounts for the fact that the average magnetization depends on the magnetic field. It is often neglected by assuming $\langle \vec{M}(\vec{B}) \rangle = \vec{M}_\text{s}$, i.e., $\mathfrak{P}(\xi) = 1$. We emphasize that $\mathfrak{P}(\xi)$ differs from $\mathfrak{L}(\xi)$ in eq. (\ref{eq:2}). The reason is that the total force on the MPs in a magnetic field gradient has two contributions: The first is the direct averaged magnetic force on MPs with average magnetization $\langle \vec{M}(\vec{B}) \rangle = \vec{M}_\text{s}\mathfrak{L}(\xi)$  (the first term in eq. (\ref{eq:8})), and the second is a thermodynamic force accounting for their drift to regions in space where their average magnetization would be energetically more favorable (the second term in eq. (\ref{eq:8})). A comparison of $\mathfrak{L}(\xi)$ and $\mathfrak{P}(\xi)$ is shown in Fig. \ref{fig2}. Both functions have a similar shape, but $\mathfrak{P}(\xi)$ reaches saturation much more rapidly and is twice as steep at the origin.
\begin{equation}\label{eq:10}
\frac{\text{d}\mathfrak{P}(0)}{\text{d}B} \;\;=\;\; 2\;\frac{\text{d}\mathfrak{L}(0)}{\text{d}B} \;\;=\;\; \frac{2}{3}\;\frac{M_\text{s}V}{k_\text{B}T} . \\
\end{equation}
Equation (\ref{eq:9}) shows that two conditions, $|\vec{B}| \neq 0$ and $\vec{\nabla}|\vec{B}| \neq 0$, must be fulfilled to enable magnetic steering of nanoparticles. Specifically, both the gradient of the field $\vec{\nabla}|\vec{B}|$ and the field-induced average magnetization $\langle \vec{M}(\vec{B}) \rangle$ must be non-zero. This complicates the characterization of magnetic systems in their ability to move MPs, because $\langle \vec{M}(\vec{B}) \rangle$ depends on the nature of the used MP. The fact that the magnetic force is governed by $\mathfrak{P}(\xi)$ rather than  $\mathfrak{L}(\xi)$, the function that describes the average magnetization of the particles, somewhat eases the problem. The average magnetization in eq. (\ref{eq:2}) will only reach a constant for $B$-values close to saturation. For iron oxide, this saturation is reached for magnetic fields with local strengths of around 1~T. However, the curve in Fig. \ref{fig2} shows that concerning the resulting magnetic force, one may approximate almost constant behavior ($\mathfrak{P}(\xi) = 1$) already at approximately 5~mT for MPs with radius of ca. $r \approx 20$~nm.\\
Nevertheless, a local magnetic field must be present to magnetize the MPs, and it must have a strong gradient to exert a force on them. 
\subsection*{Effective gradient and directionality}
Given that magnetic guiding instruments need to be evaluated for their effectiveness, a comprehensive characterization would involve a three-dimensional description or dataset of their magnetic flux density. However, comparing different designs on that ground is rather impractical, and some simple descriptors are needed. Obviously, summarizing the information of a general vector field in just a few numbers is impossible. Even if it were possible, an exact prediction of the force amplitude generated by this magnetic field is not feasible without knowledge of the characteristics of the MPs to be used in the device. This is due to the non-linearity of eq. (\ref{eq:2}) and the size-dependence of the magnetization, as illustrated in Fig. \ref{fig2}.\\
Nonetheless, it is highly desirable to extract some more meaningful quantities than, for instance, the maximal magnetic flux density and its maximal gradient. Therefore, we propose here to use a quantity that contains information on both the strength and the gradient of the field. This quantity, which we refer to as “effective gradient”, $K$, is defined by the expression:
\begin{equation}\label{eq:11}
\vec{K}(x,y,z) \;\;:=\;\; \frac{|\vec{B}(x,y,z)|}{|\vec{B}(x,y,z)|}_\text{max} \;\vec{\nabla}|\vec{B}(x,y,z)| \;\;=\;\; \frac{B}{\text{max}(B)} \;\vec{\nabla}B, \\
\end{equation}
\noindent where $|\vec{B}|_\text{max} = \text{max}(B)$ is the maximal field strength in the sample volume. Equation (\ref{eq:11}) combines amplitude and gradient of the magnitude of the magnetic field. The normalization with $|\vec{B}|_\text{max}$ has the advantage that $K$ has the units of a field gradient [T/m] and multiplication with $M_\text{s}V$ will give a rough estimate of the force that can be expected for a certain MP (assuming that $|\vec{B}|_\text{max}$ is close to the saturation field strength). As discussed earlier, this does not replace a detailed analysis, but the average, minimal, and maximal value (and possibly its standard deviation) of $|\vec{K}|$  in a specified sample volume should offer a good estimate of force distribution that can be expected in a magnetic system.\\
A second potentially important information on this magnetic vector field is the uniformity, $\vec{U}$, of the direction of the generated magnetic force. We define it as
\begin{equation}\label{eq:12}
\vec{U} \;\;:=\;\; \frac{\iiint \vec{K}\;\text{d}V}{V\,|K_{xyz}|^{(V)}_\text{max}} \;\;\approx\;\; \frac{\sum_n\vec{K}}{n \,|K_{xyz}|^V_\text{max}} \quad \text{with} \quad |K_{xyz}|^{(V)}_\text{max}\;\;:=\;\; \text{max}\left(|K_x|,\,|K_y|,\, |K_z|\right)_V , \\
\end{equation}
\noindent i.e., the integral (or, in case of discrete measurements, the sum) over all components of the effective gradient in eq. (\ref{eq:11}), normalized by the volume (or the total number of discrete entries) and $|K_{xyz}|^{(V)}_\text{max}$ which represents the maximal magnitude of all entries, $n$, of all components of $|\vec{K}|$ in that volume. For instance, $\vec{U} = (1,0,0)$ would indicate that the effective gradient points exclusively in the $x$-direction. Values less than one indicate non-uniformities in the directionality.
\begin{figure*}[t!]
\centering
\includegraphics[width=0.45\textwidth]{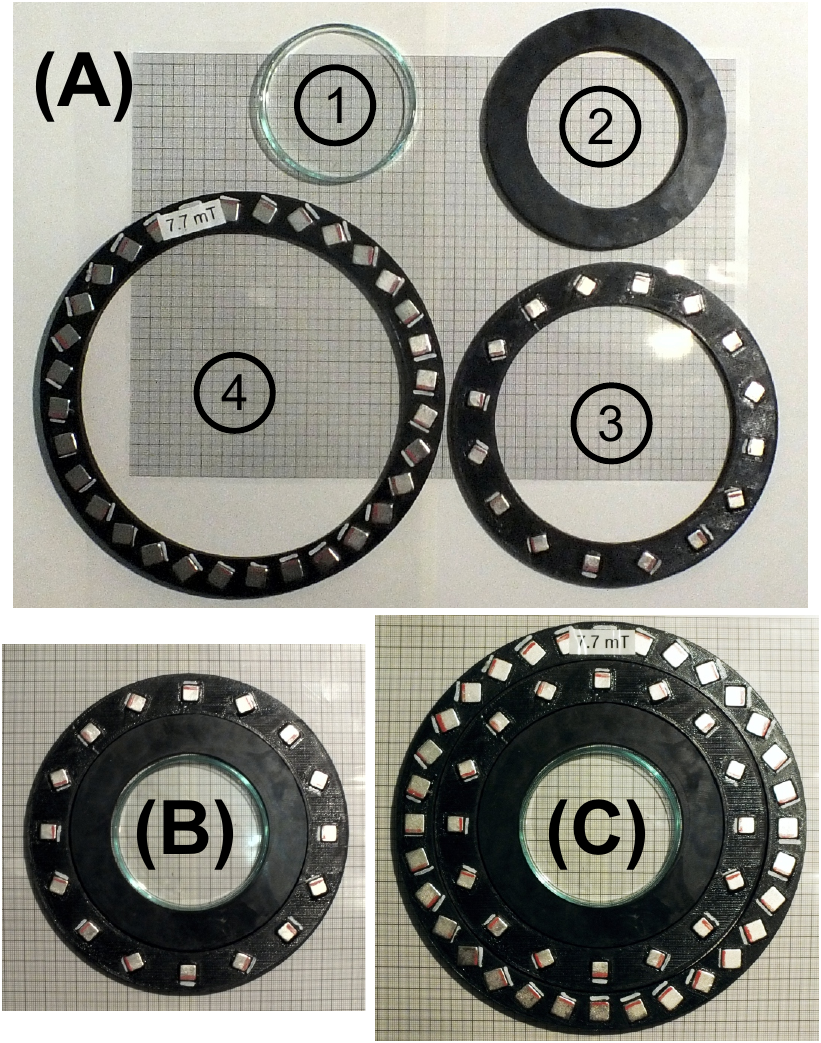}
\caption{Magnets used in the experiment: (A) photograph of all components used: (1) Petri dish (70~mm diameter), (2) 3D-printed distance ring, (3) Halbach-quadrupole, and (4) Halbach-dipole. All in front of transparent scale paper (small boxes: 1~mm\textsuperscript{2}, larger boxes. (5~mm)\textsuperscript{2}). (B) Setup of Petri dish inside the Halbach quadrupole, which provides gradients of ca. 200~mT/m. (C) Same setup as in B) but enclosed by the additional Halbach dipole, which generates a relatively homogeneous field of 7.7~mT across the sample. The combination of di- and quadrupole is named a MagGuider\cite{Baun_2017} magnet. For more details about the magnetic field, see Figs. \ref{fig4} and \ref{fig5}, and Tab. \ref{tab2}.}
\label{fig3}
\end{figure*}
\section*{Experiment}
To illustrate the key points of the prior discussion, we compare the magnetic fields of two distinct magnet designs acting on MPs like those discussed in Fig. \ref{fig2}. The first magnetic field is that of a Halbach quadrupole (cf. Fig. \ref{fig3}B) and the second is the same quadrupole combined with a Halbach dipole, forming an instrument called MagGuider\cite{Baun_2017} (cf. Fig. \ref{fig3}C). The pure quadrupole design was chosen for its provision of strong and homogeneous gradients, commonly used in experiments involving magnetically manipulated MPs or MP-loaded cells\cite{Riggio_2014,Dziob_2021,Babincova_2009}. However, the magnetic field at its center is weak (theoretically zero). Therefore, it is an ideal magnetic pattern to illustrate the difference between magnetic field gradients and force.
\begin{figure*}[t!]
\centering
\includegraphics[width=1\textwidth]{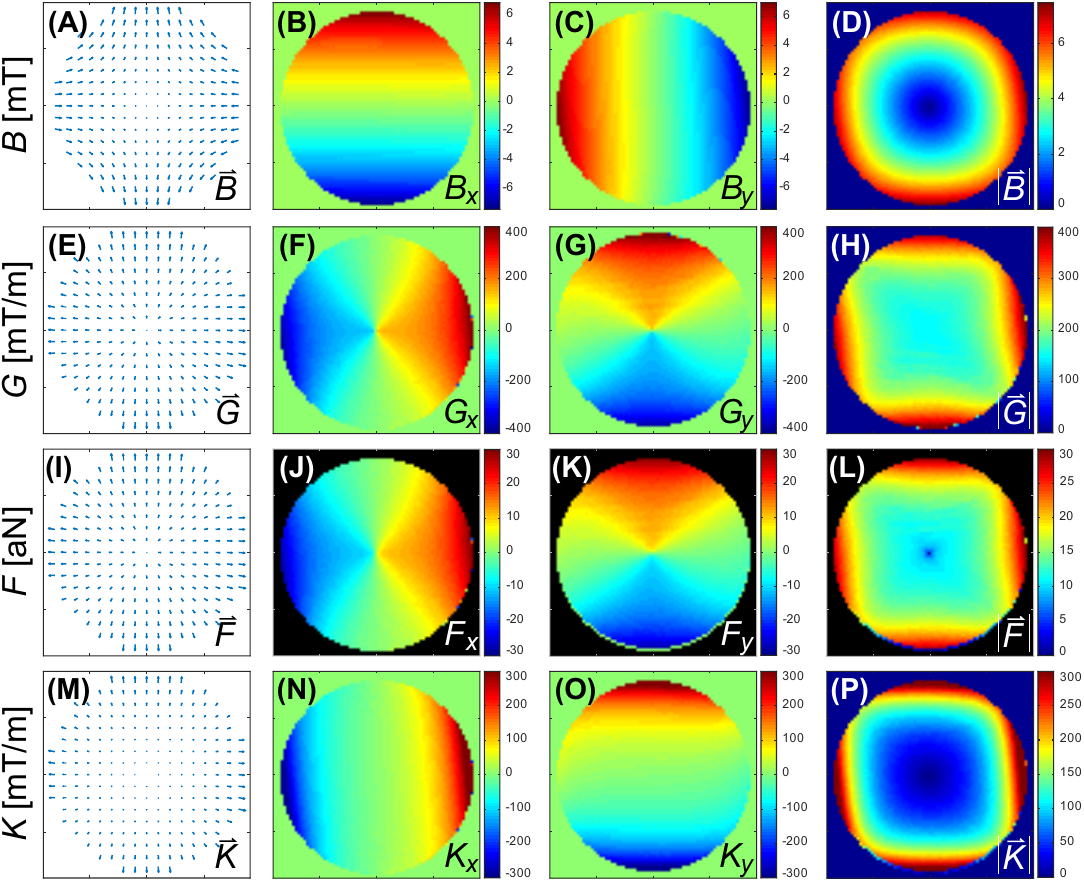}
\caption{Magnetic properties of the quadrupole in Fig. \ref{fig3}B. The magnetic field components, $B_x$ in (B) and $B_y$ in (C) were measured for a field of view of $72 \times 72\;\text{mm}^2$, for all graphs with $x$-axis in horizontal and $y$-axis in vertical direction. All the other graphs were calculated from these input data to display the features for the different entities discussed in the theoretical section. The first column shows a vector plot (direction given by arrows, amplitude by their length) of the field with reduced resolution, followed by the $x$ and $y$ component and their magnitude in the last column. The first row shows the magnetic flux density, $\vec{B}$, in mT, followed by the gradient, $\vec{G}$~[mT/m], of its magnitude in (D). The third row shows the force, $\vec{F}$,~[N], exerted on MPs with the properties of Fig. \ref{fig7} given in atto-Newton (i.e. 10\textsuperscript{-18}~N) and calculated from eq. (\ref{eq:9}). The last row shows the effective gradient, $\vec{K}$~[mT/m], as defined in eq. (\ref{eq:11}).}
\label{fig4}
\end{figure*}
The MagGuider-system is a combination of a quadrupole with a dipolar magnet, which additionally provides a homogeneous background magnetic field. It is designed to overcome the problem of low magnetic forces in certain regions. Since the homogeneous field ideally has no gradients (and only small gradients in reality), its combination with the quadrupole leads to almost identical field gradients for both systems. If the geometry is properly chosen (i.e., the condition $B > Gr$ must be fulfilled at a distance $r$ from the center, with $B$ being the strength of the homogeneous dipolar field and $G$ the constant gradient of the quadrupole), it produces a magnetic force field which is homogeneous both in amplitude and direction. The principle is explained elsewhere\cite{Bluemler_2021,Baun_2017}. Both magnets are shown in Fig. \ref{fig3}. They were designed to generate rather low magnetic fields, to illustrate the problems associated with them.\\
\begin{figure*}[t!]
\centering
\includegraphics[width=1\textwidth]{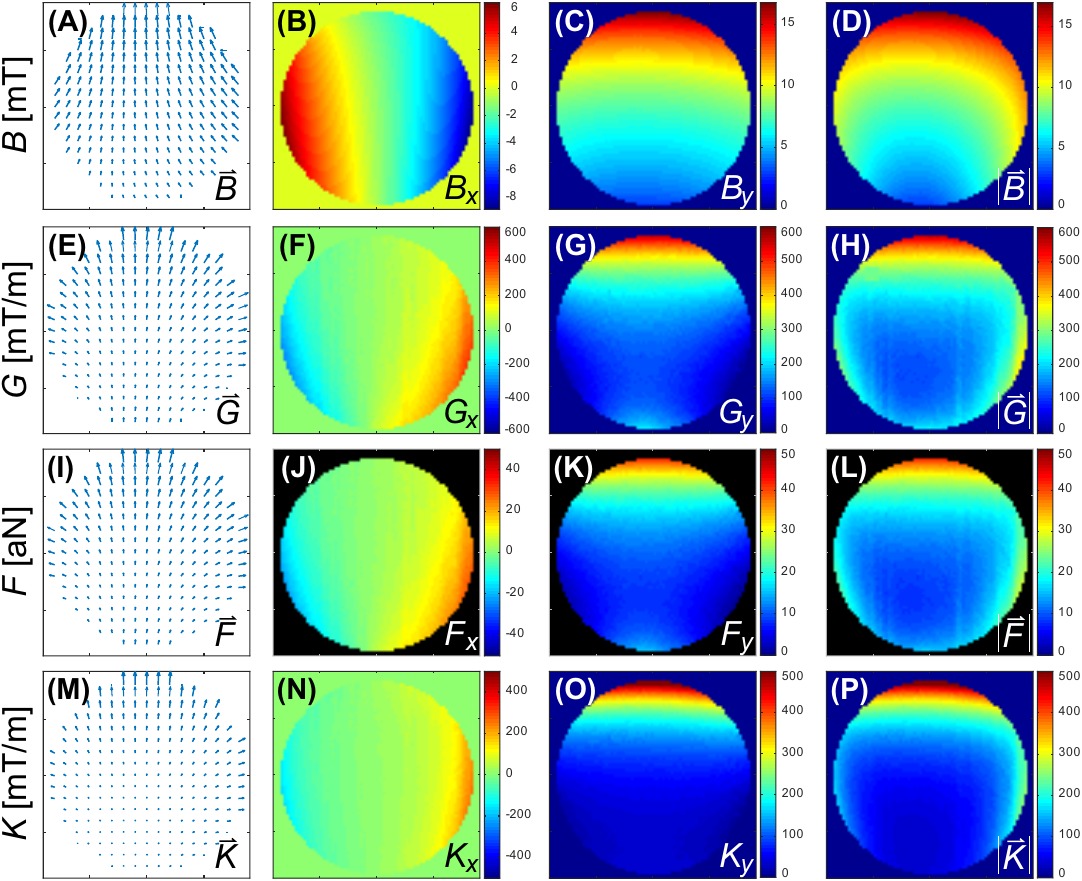}
\caption{Magnetic properties of the MagGuider system in Fig. \ref{fig3}C. Identical presentation as in Fig. \ref{fig4}.}
\label{fig5}
\end{figure*}
Figures \ref{fig4} and \ref{fig5} show the magnetic properties of both magnet systems, derived from two dimensional measurements of the $B_x$ and $B_y$ component over the size of the utilized Petri dish. In both figures, the first three columns help to understand the spatial features of the vector field and its two components, however, for the discussion the last column of the magnitude is sufficient.
In the case of the pure quadrupole system, Fig. \ref{fig4}H shows the gradient of the magnetic field magnitude in Fig. \ref{fig4}D (cf. discussion to eq. (\ref{eq:3})). This gradient is nearly constant, representing a high degree of homogeneity. However, the magnitude of the magnetic field, has radial dependence and is zero at the center as shown in Fig. \ref{fig4}D. As a consequence, the magnetization of the MPs and hence the magnetic force acting on them is very weak in the center with a strongly non-linear increase radially outwards (see Fig. \ref{fig4}L). This can be alleviated if larger MPs are used as discussed in the context of Fig. \ref{fig2}. The effective gradient is defined by eq. (\ref{eq:11}) and shown in Fig. \ref{fig4}P. This figure shows similar features to $|\vec{B}|$ in Fig. \ref{fig4}D (because its gradient is constant) but has units of a gradient. Hence, it represents the influence of the magnetic field on the force on any MP better than the field gradient (Fig. \ref{fig4}H) but overemphasizes the narrow funnel shape in Fig. \ref{fig4}L for these particular MPs.\\
The magnetic properties of the MagGuider system (Fig. \ref{fig3}C) are portrayed in Fig. \ref{fig5}. Due to the superposition of the homogeneous gradient field of the quadrupole with the homogeneous magnetic field of the dipole, the negative field, the gradient, and hence force components in the $y$-direction are eliminated. Consequently, the MPs are guided exclusively in this direction (here upwards) of the magnet system. For this particular system, the "guiding condition" ($B(\approx 8\;\text{mT}) \gg Gr(\approx 5.7\;\text{mT}))$\cite{Baun_2017} is just barely fulfilled. One can see stronger directional deviations to the sides for $x \rightarrow \pm\;35$~mm, e.g. in Fig. \ref{fig5}I. Nevertheless, it has relatively high uniformity along the $y$-direction, $\vec{U} = (-0.07,0.38)$, in particular in comparison to the quadrupole, where the uniformity is essentially zero. The presence of the homogeneous magnetic background field throughout the sample volume results in much less variation of the spatial features of G, K, and F than in the pure quadrupole system (cf. second to forth row in Fig. \ref{fig5}), as intended by the MagGuider design principle. 
\begin{figure*}[t!]
\centering
\includegraphics[width=.5\textwidth]{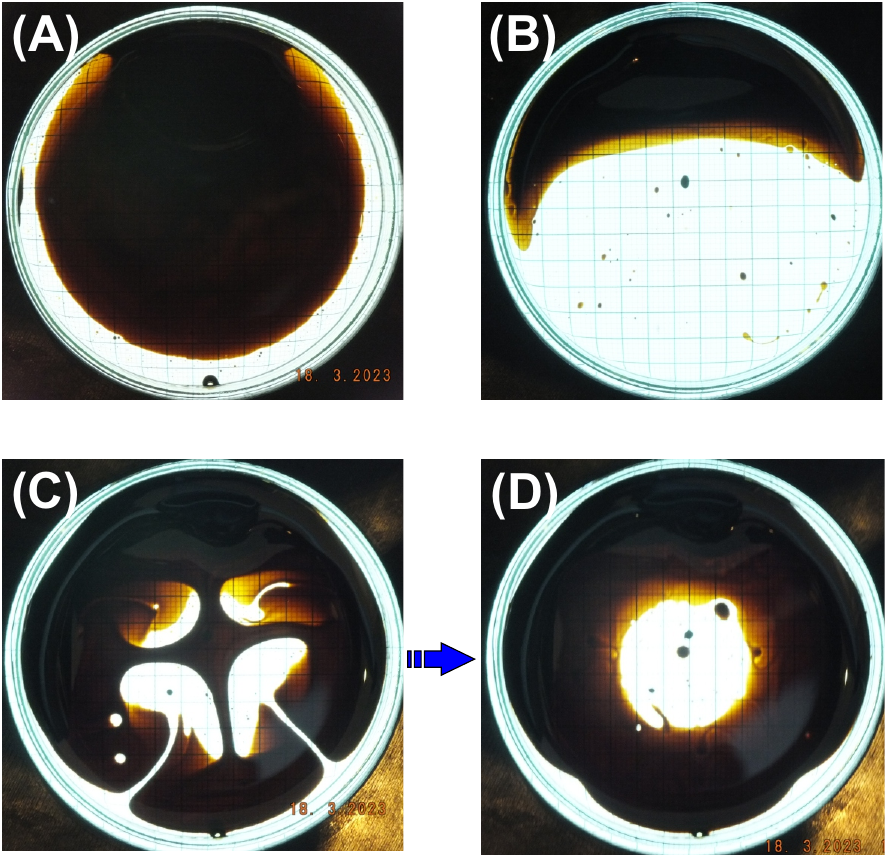}
\caption{(A) Initial state of a thin layer of oily ferrofluid of 10~nm iron particles on top of water inside the Petri dish. (B) Exposed to the magnetic field of the MagGuider magnet (cf. Figs. \ref{fig3}C and \ref{fig5}). Shown is the equilibrium state after ten seconds. (C/D) Exposed to the quadrupole only (cf. Fig. \ref{fig3}B and \ref{fig4}). (C) Photograph taken ca. 1 second after application of the magnetic field. (D) Equilibrium state after a minute.}
\label{fig6}
\end{figure*}
\begin{figure*}[t!]
\centering
\includegraphics[width=.65\textwidth]{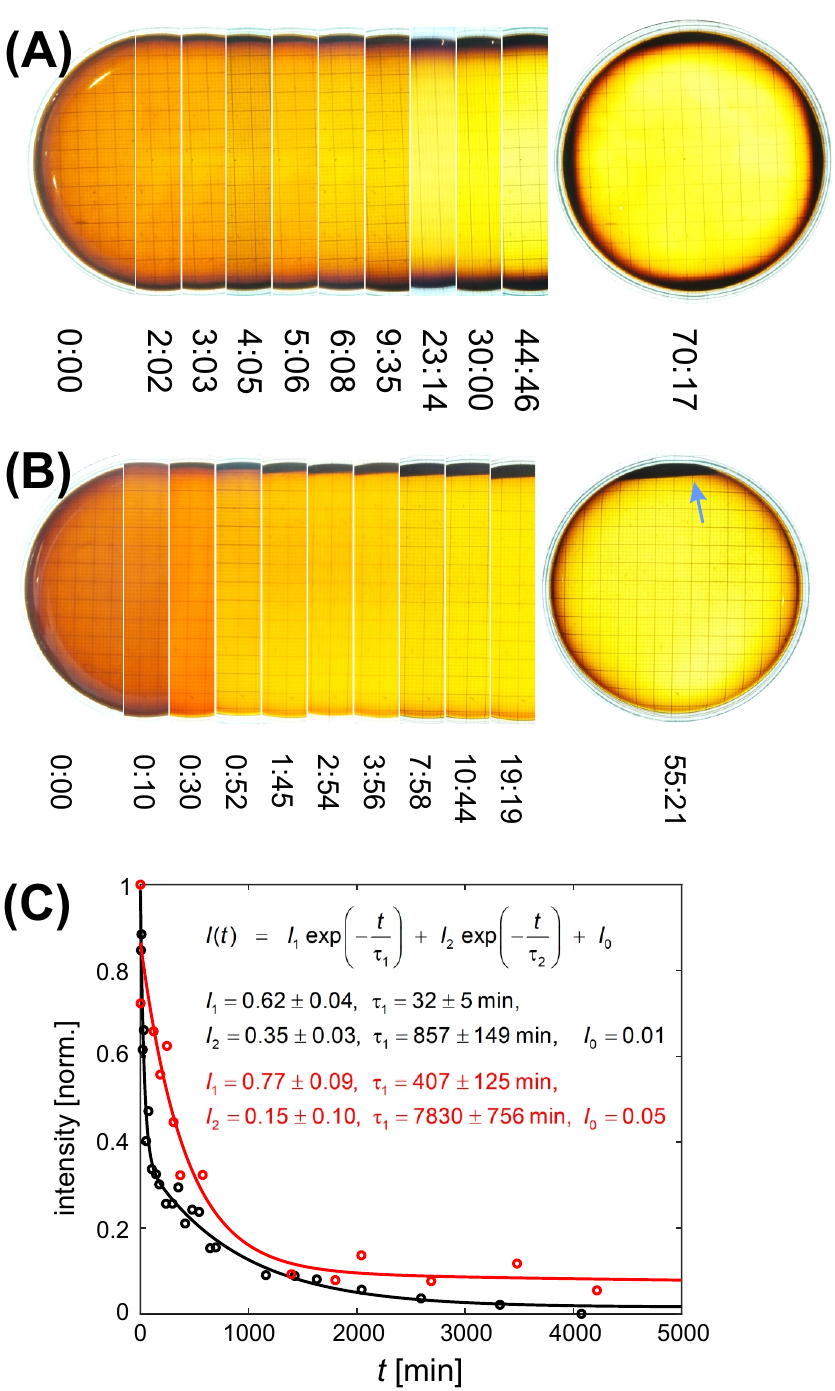}
\caption{Behavior of a dispersion of water soluble, superparamagnetic nano-particles (A) in the quadrupole, (B) in the MagGuider system. Shown are photographs at different times (hour:min) as indicated. Only central slices are concatenated except for the first image where half the Petri dish is shown. The full Petri dish is shown for the final (equilibrium) state on the right. In B) the particle concentration is indicated by a blue arrow. (C) Integration over the image intensities in a central area of ca. 2.6~cm\textsuperscript{2} and normalization to a range between 0 and 1 (black: MagGuider, red: quadrupole). Here the intensity = 1 corresponds to the initial particle concentration (cf. A/B image at 0:00) and 0 to the value obtained from an empty Petri dish. The lines correspond to a biexponential fit with the formula and the fit results given in the insert.}
\label{fig7}
\end{figure*}
Figure  \ref{fig6} provides examples of the distribution of 10~nm superparamagnetic iron-particles dispersed in light mineral oil and suspended on water at different times after exposure to the magnetic field of the quadrupole (Fig. \ref{fig3}3B and  \ref{fig4}) and the MagGuider (Figs. \ref{fig3}C and \ref{fig5}). This qualitative representation illustrates the interaction of the particles with the magnetic field of the two devices. In the MagGuider system, the initially uniformly distributed MPs in Fig.  \ref{fig6}A concentrate within one second in the direction of the magnetic force (Fig. \ref{fig6}B). Conversely, in the pure quadrupole, a fraction of particles initially moves (cf. Fig. \ref{fig6}C) towards the four poles, while a substantial portion remains in the center (“cross- shaped” feature in Fig. \ref{fig6}C), since the magnetic field (not the bare gradient) is much smaller there. In equilibrium (Fig. \ref{fig6}D), the ferrofluid concentrates at large radii, with a distribution that is mostly independent of the angle. This state is reached after approximately ten seconds. Note that, just from inspecting the forces acting on the individual MPs (cf. Tab. \ref{tab2}), one would not expect such a rapid motion. However, the velocity is not only determined by the force, but also by the hydrodynamics and the morphology. Due to the phase separation of water and ferrofluid, larger drops behave collectively (like a big particle) and slide with very low drag on top of the water surface.\\
To conduct a more quantitative analysis, water-soluble superparamagnetic nanoparticles\cite{Eberbeck_2013} were used. The results are summarized in Fig.  \ref{fig7}. Images of these particles dispersed in water were captured over a period of more than 70 hours inside the quadrupole of Fig. \ref{fig3}B. The first image on the left shows half of the Petri dish at time $t = 0$ (initial distribution), the subsequent shows a central slice at different times as indicated, and the last one the final distribution in the entire Petri dish. From the slices it can be recognized that the initial brownish color of the dispersion becomes brighter with time as more and more particles are dragged towards the walls. This happens much faster in the case of the MagGuider systems, as shown in Fig. \ref{fig7}B (same mode of presentation). Please note the very different time scale in both figures. In the final state the particles end up at all walls in the case of the pure quadrupole, and at the top rim (indicated by the blue arrow) in the case of the MagGuider as already shown in Fig. \ref{fig6}.\\
In order to analyze this more quantitatively, the intensity of a central region of each image was integrated and normalized with respect to an image of the empty Petri dish and the intensity of the initial particle concentration. Fig. \ref{fig7}C shows the results, which were fitted by a double exponential decay. The fast decay is attributed to the drag exerted on the nanoparticles by magnetic forces. The slower decay reflects the motion of some remaining slower particles, which might be slowed down by interactions with the bottom of the dish, or -– in the case of the quadrupole, an unfavorable trajectory: Particles that initially stay in the low field in the center first have to diffuse into a regions with stronger magnetic forces (diffusing water needs ca. 2200~min or 37~h to cover a distance of 35~mm). Both the time constant of the initial fast intensity decay and the subsequent slower decay are nearly an order of magnitude smaller for the MagGuider system, even though the gradients are almost identical in both systems (cf. Tab. \ref{tab2}).\\
In Tab. \ref{tab2}, we attempt to characterize the relevant properties of both magnetic systems by the descriptive numbers discussed above, namely, $|\vec{B}| = B$, $|\vec{G}| = G$, and $|\vec{K}| = K$, and test their use for predicting the magnetic force on the two different particles used in Figs. \ref{fig6} and \ref{fig7}. The most obvious discrepancy arises from the much lower force measured for the smaller particles ($r = 5$~nm) used in Fig. \ref{fig6}. Here the predictions via $M_\text{s}V\;|\vec{G}|$ or $M_\text{s}V\;|\vec{K}|$  deviate by one or two orders of magnitude from the calculated force for both magnet systems. This is because the magnetic field in both systems is too low to fully magnetize the MPs, and resulting in $\overline{\mathfrak{P}}(\xi) = 10^{-3}$ for the quadrupole and $\overline{\mathfrak{P}}(\xi) = 0.02$ for the MagGuider (where $\overline{\mathfrak{P}}(\xi)$ is the averaged value in Tab. \ref{tab2}). This illustrates the importance of knowing the local magnetic fields. The predictions using $M_\text{s}V\;|\vec{K}|$ instead of $M_\text{s}V\;|\vec{G}|$ are slightly better, but still differ from $|\vec{F}| = F$, in particular when looking at the minimum value. The same holds for the larger particles ($\overline{\mathfrak{P}}(\xi) \approx 1$), where the value of $M_\text{s}V\;|\vec{G}|$ is identical with the force for the MagGuider, but again overestimate the minimal force by an order of magnitude for the quadrupole. The best description is realized by giving the range of all three quantities $B$, $G$, and $K$, the latter because it connects the local information on the other two. Without this information (e.g., by only giving the gradient strengths of the systems, which are close to identical for both of them), the very different behavior of MPs as in Fig. \ref{fig7}C can neither be explained nor predicted.\\
\begin{table}[h!]
\centering
\renewcommand{\arraystretch}{1.5}
\begin{tabular}{c l|c|c|c|c}
   magnet & quantity & average & std. dev. & max. & min.\\
   \hline\hline
   \multirow{9.3}{*}{\begin{sideways} quadrupole \end{sideways}}
   &$|\vec{B}|$ [mT]& 2.08 & 0.77 & 3.34 & 0.01*\\
   \cline{2-6}
   &\cellcolor{lightgray}$|\vec{G}|$ [mT/m]& \cellcolor{lightgray} 168.1 &\cellcolor{lightgray} 14.9 &\cellcolor{lightgray} 207.2 &\cellcolor{lightgray} 8.4\\
   \cline{2-6}
   &$|\vec{K}|$ [mT/m]& 107.7 & 46.6 & 207.2 & 0.02*\\
   \cline{2-6}
   &\cellcolor{lightgray}$|\vec{F}|\;\; [10^{-22}\;\text{N}]\; (r = 5\;\text{nm})$&\cellcolor{lightgray} 0.16 &\cellcolor{lightgray} 0.07 &\cellcolor{lightgray} 0.31 &\cellcolor{lightgray} $3\,\cdot\,10^{-5}\text{*}$\\
   \cline{2-6}
   &\cellcolor{lightgray}$M_\text{s}V|\vec{G}|\;\; [10^{-22}\;\text{N}]\; (r = 5 \;\text{nm})$&\cellcolor{lightgray} 28.1 &\cellcolor{lightgray} 2.5 &\cellcolor{lightgray} 34.7 &\cellcolor{lightgray} 1.4\\
   \cline{2-6}
   &\cellcolor{lightgray}$M_\text{s}V|\vec{K}|\;\; [10^{-22}\;\text{N}]\; (r = 5\; \text{nm})$&\cellcolor{lightgray} 18.0 &\cellcolor{lightgray} 7.8 &\cellcolor{lightgray} 34.7 &\cellcolor{lightgray} 0.003*\\
   \cline{2-6}
   &$|\vec{F}|\;\; [10^{-18}\;\text{N}]\; (r = 65\;\text{nm})$& 23.2 & 2.1 & 28.6 & 0.2*\\
   \cline{2-6}
   &$M_\text{s}V|\vec{G}|\;\; [10^{-18}\;\text{N}]\; (r = 65\;\text{nm})$& 23.2 & 2.1 & 28.6 & 1.1\\
   \cline{2-6}
   &$M_\text{s}V|\vec{K}|\;\; [10^{-18}\;\text{N}]\; (r = 65\;\text{nm})$& 14.9 & 6.4 & 28.6 & 0.003*\\
   \Xhline{4\arrayrulewidth}
   \multirow{9.2}{*}{\begin{sideways} MagGuider \end{sideways}} 
   &$|\vec{B}|$ [mT]& 7.96 & 1.54 & 11.59 & 5.09\\
   \cline{2-6}
   &\cellcolor{lightgray}$|\vec{G}|$ [mT/m]& \cellcolor{lightgray} 163.6 &\cellcolor{lightgray} 39.9 &\cellcolor{lightgray} 287.2 &\cellcolor{lightgray} 108.9\\
   \cline{2-6}
   &$|\vec{K}|$ [mT/m]& 117.6 & 52.1 & 287.2 & 49.5\\
   \cline{2-6}
   &\cellcolor{lightgray}$|\vec{F}|\;\; [10^{-22}\;\text{N}]\; (r = 5\;\text{nm})$&\cellcolor{lightgray} 0.62 &\cellcolor{lightgray} 0.27 &\cellcolor{lightgray} 1.50 &\cellcolor{lightgray} 0.26\\
   \cline{2-6}
   &\cellcolor{lightgray}$M_\text{s}V|\vec{G}|\;\; [10^{-22}\;\text{N}]\; (r = 5\;\text{nm})$&\cellcolor{lightgray} 27.4 &\cellcolor{lightgray} 6.7 &\cellcolor{lightgray} 48.1 &\cellcolor{lightgray} 18.2\\
   \cline{2-6}
   &\cellcolor{lightgray}$M_\text{s}V|\vec{K}|\;\; [10^{-22}\;\text{N}]\; (r = 5\;\text{nm})$&\cellcolor{lightgray} 19.7 &\cellcolor{lightgray} 8.7 &\cellcolor{lightgray} 48.1 &\cellcolor{lightgray} 8.3\\
   \cline{2-6}
   &$|\vec{F}|\;\; [10^{-18}\;\text{N}]\; (r = 65\;\text{nm})$& 22.6 & 5.5 & 39.6 & 15.0\\
   \cline{2-6}
   &$M_\text{s}V|\vec{G}|\;\; [10^{-18}\;\text{N}]\; (r = 65\;\text{nm})$& 22.6 & 5.5 & 39.6 & 15.0\\
   \cline{2-6}
   &$M_\text{s}V|\vec{K}|\;\; [10^{-18}\;\text{N}]\; (r = 65\;\text{nm})$& 16.2 & 7.2 & 39.6 & 6.8\\
   \hline
\end{tabular}
\caption{Characterization of the two magnet systems (quadrupole and MagGuider) by a few descriptive values representing the full distributions shown in Figs. \ref{fig4} and \ref{fig5}. From those the arithmetic average, standard deviation, maximal and minimal value are determined over a central circular area of 40~mm in diameter. This was done for magnitude of the magnetic flux density $|\vec{B}|$, its gradient, $|\vec{G}|$, effective gradient $|\vec{K}|$. For the two particles used in Figs. \ref{fig6} and \ref{fig7} the magnetic force $|\vec{B}|$ (using eq. (\ref{eq:9})) and two simplified (scalar) values ($M_\text{s}V\,|\vec{G}|$ and  $M_\text{s}V\,|\vec{K}|$ ) are compared. The directionality is $\vec{U} = (2, 0.3)\,\cdot\,10^{-3}$ for the quadrupole and $\vec{U} = (-0.07, 0.38)$ for the MagGuider magnet. (*the value of the minimum must be zero, because $|\vec{B}(0,0)| = 0$. The non-zero values are due to insufficient discretization of the field and/or imprecise measurements.)}
\label{tab2}
\end{table}
\section*{Discussion}
In the application of magnetic systems for magnetic guiding, it is crucial to have a standardized characterization and description. This not only aids in comprehending, interpreting, comparing, and replicating results, but also provides a more comprehensive assessment than relying on the maximal field gradient alone. This is because both the thermally averaged magnetization and the field gradient contribute to the overall force. While the average magnetization is an inherent material property of the used MPs, it is also dependent on the magnetic field strength. Consequently, both the magnitude of the magnetic flux and its gradient are essential in evaluating a device’s capacity to exert a force on MPs. This is illustrated by two magnet systems, that possess similar gradient strengths but exhibit a tenfold difference in their ability to move MPs. 
Therefore, we propose to introduce a different quantity termed ‘effective gradient’, which describes the joint spatial characteristics of $|\vec{B}|$ and $\vec{\nabla}|\vec{B}|$. The spatial distribution of this "effective gradient" can provide a rough estimate of the generated magnetic force and can be approximated using typical descriptors of distributions like average, standard deviation, minimum, and maximum over the volume of interest. Additionally, providing information about the magnetic field and gradient separately can further enhance characterization. Furthermore, the orientational homogeneity of the force can be described by a directionality vector. We hope that these suggestions will facilitate a better comparison of instruments designed for magnetic guiding.\\
Although magnetic forces acting on single nanoparticles are in the sub-attonewton regime and might appear extremely low, they are sufficient to induce aggregate formation depending on the coating and concentrations of MPs due the attraction via dipole forces. The force on such an aggregate is then the multiple of the force on a single MP, scaled by the number of particles in the aggregate (cf. the ferrofluid drops in Fig. \ref{fig6}).\\
This underscores that even knowing the magnetic force alone is insufficient to fully describe the ability of a system for magnetic guiding, because it is the motion of the MPs, which is of interest. The relevant quantities are the velocity and acceleration of the MPs. To determine them becomes a hydrodynamical problem, which can be described by Stokes’ law for MPs or MP aggregates in liquids (for details and the hydrodynamics of aggregates see\cite{Bluemler_2021}. In addition, Hooke’s law applies in elastic, isotropic solids, and theories of viscoelasticity come into play in complex fluids like polymer or protein solutions. Consequently, local factors such as MP concentration, and rheological characteristics of the surrounding medium such as viscosities or moduli are needed to predict the particle motion from the force field. In the complex and heterogeneous environment of biological tissues, simulations, ideally assisted by experiments, are required to predict the forces needed to induce the intended motion.\\
\section*{Materials and Methods}
The magnets depicted in Fig. \ref{fig3} are constructed using cubic permanent neodymium magnets (Nd\textsubscript{2}Fe\textsubscript{14}B) affixed into 3D printed supports (polylactic acid). Specifically, the quadrupole is assembled from 16 cubic magnets with 8~mm side length (N45, $B_\text{R} = 1.35$~T) obtained from magnets4you GmbH, 97816 Lohr, Germany. Their centers are mounted in a circle with a radius $r = 67.5$~mm.\\
The dipole consists of 32 magnets (10~mm cubes, N45, $B_\text{R} = 1.35$~T) purchased from dogeo GmbH, 55444 Waldlaubersheim, Germany, and mounted in a circular arrangement with $r = 87$~mm.\\
The magnetic field components in Figs. \ref{fig4} and \ref{fig5} were measured with a Hall-probe sensor (HMNT-4E04-VR, Lake Shore Cryotronics Inc., Westerville, OH) mounted on a computer-controlled 3D linear table (OWIS GmbH, Staufen im Breisgau, Germany) and controlled by custom-written software. A circular area with a radius of 36~mm was scanned with a step width of 1~mm in $x$ and $y$ direction.\\
The MPs used in the qualitative display shown in Fig. \ref{fig6} are 10~nm are iron particles (EFH1, $M_\text{s} = 32$ kA/m) dispersed in light mineral oil, acquired from FerroTec, Santa Clara, CA.\\
The MPs used in the experiments for Fig. \ref{fig7} are 130~nm nanomag-D with plain surfaces (product no. 09-00-132) from micromod, Rostock, Germany. They are multicore iron oxide particles ($M_\text{s} = 120$~kA/m) consisting of several smaller magnetic particles\cite{Eberbeck_2013}.\\ 

\section*{Acknowledgements}
We wish to thank Rudolf Zentel (University of Mainz) and Matthias Barz (University of Leiden) for inspiring this work. Rolf Heumann (University of Bochum) has to be acknowledged for helpful discussions and suggestions regarding this manuscript.\\
This work has been supported by the Deutsche Forschungsgemeinschaft funded Collaborative Research Center \textit{“Nanodimensional polymer therapeutics for tumor therapy”} – SFB 1066 , (DFG grant number 213555243). Further, we thank JSPS-KAKENHI (grant number 22K06430, F.R.) for the support.
\section*{Author contributions statement}
P.B. build the magnets and performed the experiments, otherwise all authors equally contributed to the manuscript and its revisions. 
\end{document}